\documentclass{aa}  

\usepackage{xcolor}
\usepackage{graphicx}
\usepackage{txfonts}
\usepackage{hyperref}
\usepackage[utf8]{inputenc}
\begin{document}

   \title{Peculiar Hydrogen-deficient Carbon Stars: Strontium-Rich Stars and the s-Process}

%   \subtitle{I. Overviewing the $\kappa$-mechanism}

   \author{Courtney L. Crawford
          \inst{1}
          \and
          Patrick Tisserand\inst{2}
          \and
          Geoffrey C. Clayton\inst{1}
          \and
          Bradley Munson\inst{1}
          }

   \institute{Dept. of Physics \& Astronomy, Louisiana State University, Baton Rouge, LA, 70803, USA\\
              \email{ccour14@lsu.edu}
         \and
             Sorbonne Universit\'es, UPMC Univ. Paris 6 et CNRS, UMR 7095, Institut d’Astrophysique de Paris, IAP, 75014 Paris, France\\
             }

   \date{Received ---; ---}
 
  \abstract
   {R Coronae Borealis (RCB) variables and their non-variable counterparts, the dustless Hydrogen-Deficient Carbon (dLHdC) stars have been known to exhibit enhanced  \textit{s}-processed material on their surfaces, especially Sr, Y, and Ba. No comprehensive work has been done to explore the \textit{s}-process in these types of stars, however one particular RCB star, U Aqr, has been under scrutiny for its extraordinary Sr enhancement.}
   {We aim to identify RCB and dLHdC stars that have significantly enhanced Sr abundances, such as U Aqr, and use stellar evolution models to begin to estimate the type of neutron exposure that occurs in a typical HdC star.}
   {We compare the strength of the Sr II 4077 $\AA$ spectral line to Ca II H  to identify the new subclass of Sr-rich HdCs. We additionally use the structural and abundance information from existing RCB {\sc MESA} models to calculate the neutron exposure parameter, $\tau$.}
   {We identify six stars in the Sr-rich class. Two are RCBs, and four are dLHdCs. We additionally find that the preferred RCB {\sc MESA} model has a neutron exposure $\tau$ $\simeq$ 0.1 mb$^{-1}$, which is lower than the estimated $\tau$ between 0.15 and 0.6 mb$^{-1}$ for the Sr-rich star U Aqr found in the literature. We find trends in the neutron exposure corresponding to He-burning shell temperature, metallicity, and assumed \textit{s}-processing site.}
   {We have found a sub-class of 6 HdCs known as the Sr-rich class, which tend to lie in the halo, outside the typical distribution of RCBs and dLHdCs. We find that dLHdC stars are more likely to be Sr-rich than RCBs, with an occurrence rate of $\sim$13\% for dLHdCs and $\sim$2\% for RCBs. This is one of the first potential spectroscopic differences between RCBs and dLHdCs, along with dLHdCs having stronger surface abundances of $^{18}$O. We additionally find neutron exposure trends in our RCB models that will aide in understanding the interplay between model parameters and surface \textit{s}-process elements.}


   \maketitle
%
%-------------------------------------------------------------------

\section{Introduction}

The rare supergiant class of variables known as the R Coronae Borealis (RCB) stars and their non-variable counterparts, the dustless Hydrogen-deficient Carbon (dLHdC) stars, provide a wealth of information on their unique stellar evolution. These two subsets of stars form the overarching class of Hydrogen-deficient Carbon (HdC) stars.\footnote{dLHdCs have historically been referred to simply as HdC stars. However, following the nomenclature of \citet[][]{Tisserand2022_submitted}, we refer to them as dLHdCs to ease the confusion between them and the parent class of HdC stars, which includes both the RCB stars and the dLHdCs.} While spanning a higher temperature range (5000-8000 K) than the carbon star population (2500-6000 K, \citealp{Keenan1993_carbonclass}), these stars are spectroscopically similar to other carbon stars, except with very weak or undetectable H lines and CH bands, a severe depletion in $^{13}$C, and strong $^{12}$C$^{18}$O bands in the infrared \citep{Clayton1996_review,Clayton2007_O18,Clayton2012_review}. These stars also show enhanced \textit{s}-process elements, for which the primary neutron source is the $^{13}$C($\alpha$,n)$^{16}$O reaction \citep{Asplund2000}.

The origin of RCB stars was originally theorized with two possible paths: a post-AGB star undergoing a very late thermal pulse (often called a final He-shell flash or the FF model), and the merger of a Carbon/Oxygen (CO-) and a Helium (He-) white dwarf (WD) \citep{Fujimoto1977_ffmodel,Webbink1984_ddmodel}. Recently, the former evolutionary path has fallen out of favor for multiple reasons. Namely, there are three stars known to have undergone a very late thermal pulse: Sakurai's Object, V605 Aql, and FG Sge, and while these stars do spectroscopically resemble RCB stars, they do not reflect important aspects of the RCB spectrum such as enhanced $^{18}$O abundance and weak $^{13}$C \citep{Clayton2012_review}. Additionally, their lifetimes are too short to explain the population size of known RCB stars. Recent RCB models originating from WD mergers have been successful at replicating the $^{16}$O/$^{18}$O and $^{12}$C/$^{13}$C ratios of RCB stars, as well as the average radius, mass and temperatures for these stars \citep{Longland2011,Menon2013,Menon2018,Zhang2014,Schwab2019,Lauer2019,Crawford2020_mesa,Munson2021}.

The \textit{s}-process abundances in HdC stars are often compared to the "strong" and "weak" components of the Solar System \textit{s}-process abundances. The former component, which makes up the majority of these abundances, is understood as enrichment from AGB stars which have undergone several thermal pulses and therefore a series of decaying neutron exposures, the sum of which can be modeled as a weighted exponential neutron exposure \citep{Beer1989_sprocess}. The latter component, on the other hand, can be understood as a weaker single neutron exposure, the source of which is not well understood \citep{Beer1989_sprocess}. Utilizing the models created by \citet{Crawford2020_mesa}, in Section~\ref{sec:sprocess} we estimate a probable neutron exposure for a typical HdC star and compare to those that create the Solar System \textit{s}-process abundances.

\begin{figure}
\centering
\includegraphics[width=\columnwidth]{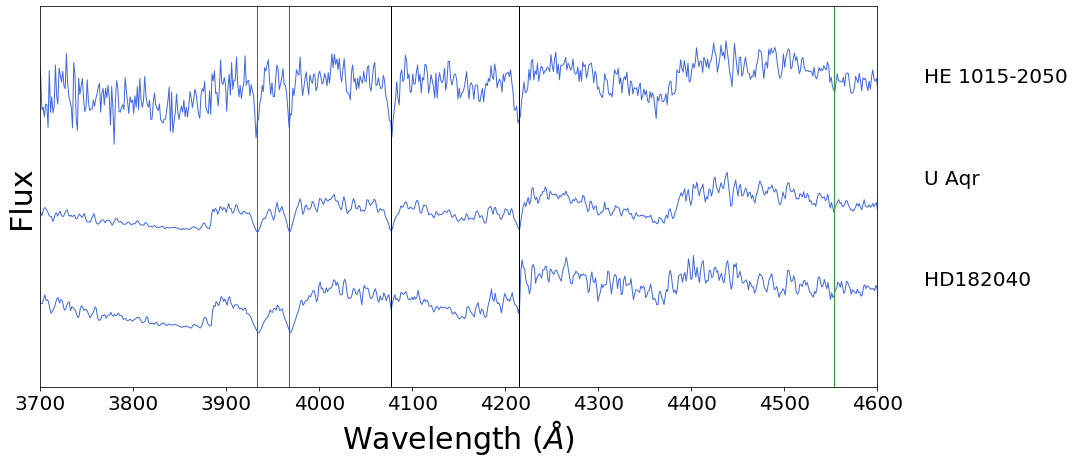}
\caption{The blue region of the spectrum for U Aqr and HE 1015-2050 with a non-Sr-rich comparison star of similar temperature, HD 182040. Ca II H and K are denoted by vertical red lines, Sr II 4077 and 4215 by vertical black lines, and Ba II 4554 by a vertical green line.}
          \label{fig:compstar}%
\end{figure}

One particularly unusual RCB star, U Aqr, has been recognized for its unique \textit{s}-process element distribution. In particular, the light \textit{s}-process elements such as Sr, Y, and Zr are greatly enhanced in comparison to heavier \textit{s}-process elements such as Ba (see Figure~\ref{fig:compstar}) \citep{Bond1979_uaqr,Malaney1985_uaqr,Vanture1999_uaqr}. Bond referred to U Aqr as "the star with the strongest known Strontium lines," as the blue Sr II lines at 4077\text{\AA} and 4215\text{\AA} are comparable in strength to Ca H and K.
\citet{Vanture1999_uaqr} suggest that U Aqr's abundances most resemble the weak single neutron exposure component. 
\citet{Goswami2013_he1015sr} also point out that the dLHdC star HE 1015-2050 has similarly strong Sr II lines in the blue region of the spectrum. 
In Section~\ref{sec:line_analysis} we present a total of six HdCs with similarly enhanced \textit{s}-process distributions: U Aqr, HE 1015-2050, EROS2-LMC-RCB-3, A249, A166, and C539. In Section~\ref{sec:discussion} we summarize the known properties of these stars and hypothesize why their neutron exposures could be different from other HdC stars.

%%%%%%%%%%%%%%%%%%%%%%%%%%%%%%%%%%%%%%%%%%%%%%%%%%%%%%%%%%%%%%%%%%%%%%%%%
\section{Observations}
\label{sec:observations}

Currently there are 157 RCB stars and 32 dLHdC stars known in the Milky Way and the Magellanic Clouds \citep{Tisserand2020_plethora,Tisserand2022_submitted}. We have obtained optical spectra for 144 of these stars which are being used for HdC star spectral classification (Crawford et al. 2022, in prep.). These spectra were obtained with the Wide Field Spectrograph (WiFeS) \citep{Dopita2007_wifes} mounted on the 2.3 m telescope of the Australian National University at Siding Spring Observatory (SSO). Specific details on the spectral acquisition can be found in \citet{Tisserand2020_plethora} and \citep[][]{Tisserand2022_submitted}. The spectra span the wavelengths 3400\text{\AA} to 9600\text{\AA} with a two-pixel resolution of about 2\text{\AA}. See Table~\ref{tab:obs} for details of the observations of stars with exceptional \textit{s}-process abundances. 

The RCB and dLHdC star spectra must be dereddened as the stars are predominantly found at large distances. 
%However, many of these stars are in the foreground of a significant amount of dust. 
We first searched \citet{Green2019_extinction} for an A$_v$ value. If there were no data there, we then adopted the A$_v$ value from \citet{S_and_F2011_extinction}. We applied this extinction correction using an average R(V) = 3.1 and CCM dust \citep{CCM89_dust}.

\begin{table}
\caption{Observations of Sr-rich Stars}             % title of Table
\label{tab:obs}      % is used to refer this table in the text
\centering                          % used for centering table
\begin{tabular}{c c}        
\hline                % inserts double horizontal lines
Name & Observation Date \\    % table heading 
\hline                        % inserts single horizontal line
   U Aqr & 15 July 2010 \\      % inserting body of the table
   HE 1015-2050 & 7 June 2012 \\
   EROS2-LMC-RCB-3 & 18 April 2008 \\
   A249 & 14 April 2021 \\
   A166 & 21 May 2021 \\ 
   C539 & 24 September 2021 \\
\hline                                   %inserts single line
\end{tabular}
\end{table}

%%%%%%%%%%%%%%%%%%%%%%%%%%%%%%%%%%%%%%%%%%%%%%%%%%%%%%%%%%%%%%%%%%%%%%%%%

\section{Spectral Line Analysis}
\label{sec:line_analysis}

Equivalent widths (EWs) were measured for the available s-process lines.
The spectra of HdC stars are full of weak atomic lines and, for the cooler stars, molecular bands. 
Therefore, the spectral continuum is not directly observable at intermediate resolution and we must rely instead on pseudo-continua and, by extension, pseudo-equivalent widths (hereafter pseudo-EW) of the measured lines. We examine all the lines listed in Table~\ref{tab:lines}. Many of the spectral lines, observed at this intermediate resolution, especially those for weak-lined elements such as Fe, are blended with other lines. Therefore, we are unable to estimate on the intrinsic abundances of HdCs, opting only to measure the pseudo-EW of the stronger lines in the spectrum. High resolution abundance analyses have been done for a few individual HdC stars \citep[e.g.,][]{Asplund2000}.

\begin{figure}
\centering
\includegraphics[width=\columnwidth]{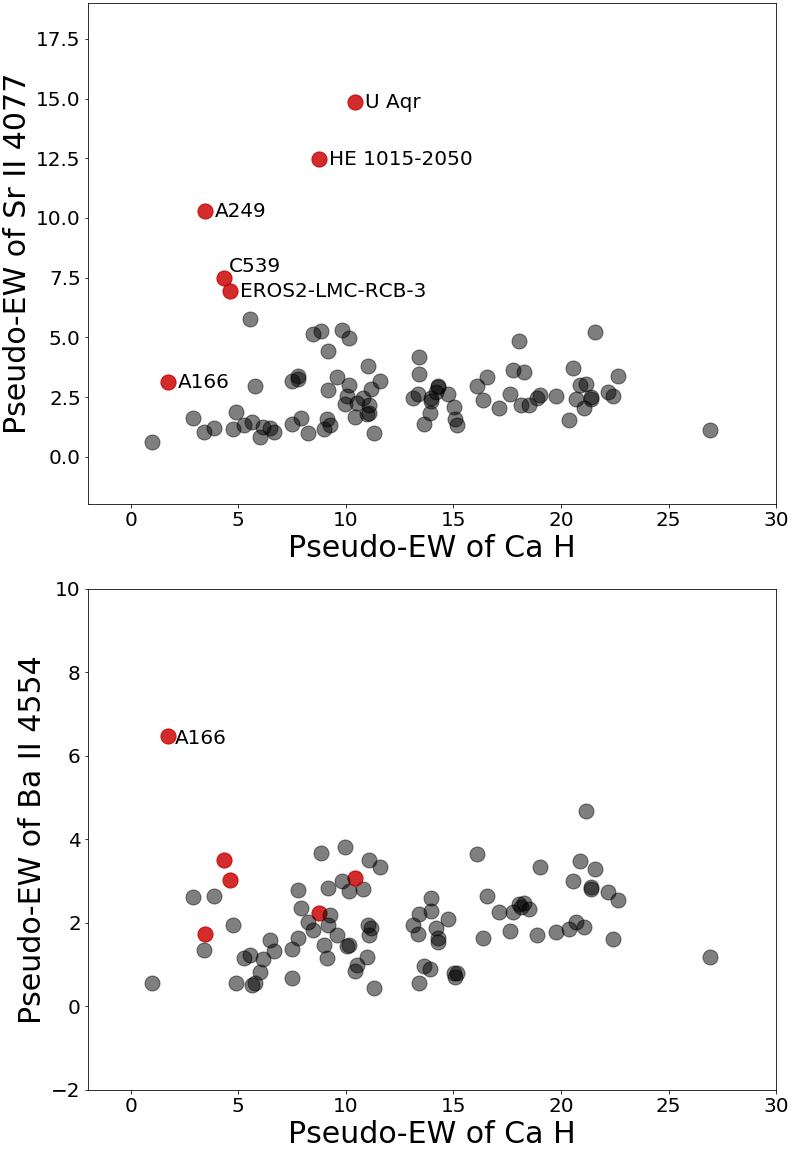}
\caption{The pseudo-EW of Sr II 4077 vs Ca II H in the upper panel and Ba II 4554 vs Ca II H in the lower panel. The Sr-rich stars are plotted in red and labeled. In the lower panel, only A166 is labeled so as not to obstruct the rest of the data.}
          \label{fig:sr_v_ca}%
\end{figure}

\begin{figure*}
\centering
\includegraphics[width=7in]{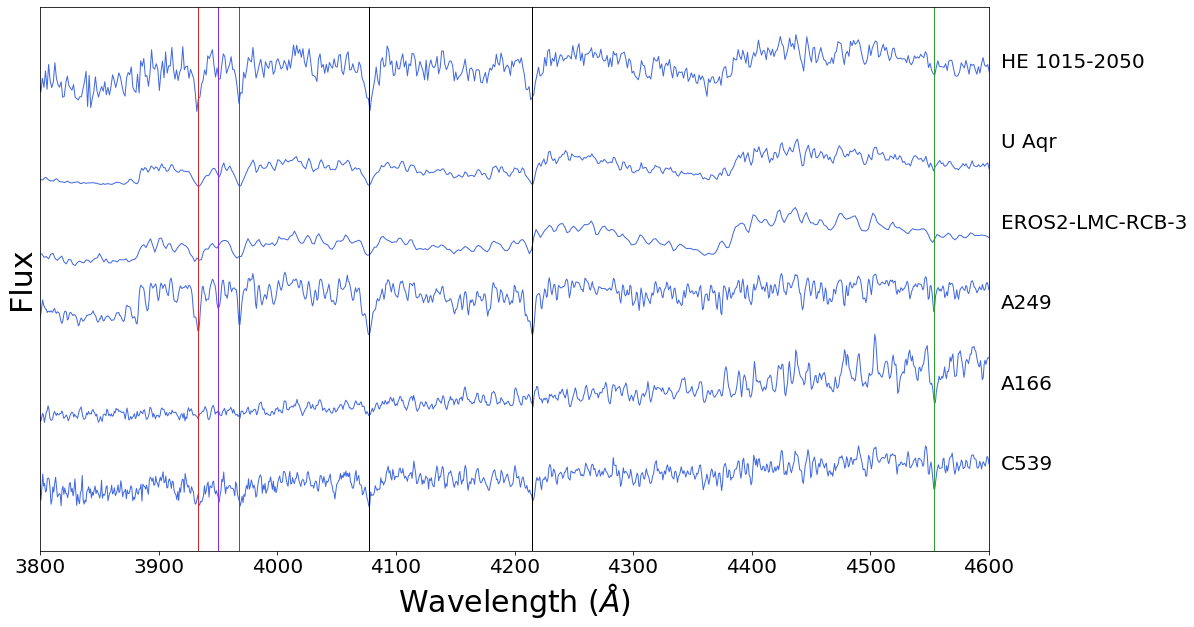}
\caption{The blue region of the spectrum for the six Sr-rich stars. Ca II H and K are denoted by vertical red lines, Sr II 4077 and 4215 by vertical black lines, Y II by a vertical purple line, and Ba II 4554 by a vertical green line.}
          \label{fig:srrichstars}%
\end{figure*}

For each HdC star, we compare the pseudo-EWs of the Ca II H \& K lines to those of the nearby Sr II lines (4077 and 4215 $\AA$). We calculate the ratio of Sr II 4077 to Ca II H, as the Sr II 4215 line is blended with a CN band head, and the Ca II K line is close to the CN bandhead at $\sim$3890 $\AA$. The stars where the Ca II H/Sr II 4077 ratio is less than 1.5, i.e., where the Sr II pseudo-EW is comparable or larger than that of the Ca II H lines, are a special class of HdC star we denote as "Sr-rich". By comparison, the average value of this ratio for HdC stars that are not rich in Sr is $\sim$5.5. Comparing the pseudo-EWs of Sr II 4077 and Ca II H for all HdC stars allows for clear identification of the six Sr-rich stars (see Figure~\ref{fig:sr_v_ca}): U Aqr, EROS2-LMC-RCB-3, HE 1015-2050, A249, A166, and C539. U Aqr and EROS2-LMC-RCB-3 are known RCB stars, HE 1015-2050 is a known dLHdC and the latter three are new dLHdC stars, discovered by \citet[][]{Tisserand2022_submitted}. Their blue spectra are shown in Figure~\ref{fig:srrichstars}. 

We chose Ca II H \& K for this analysis as they are usually some of the strongest lines in the spectrum. Upon some testing, we could not identify any trends with the pseudo-EWs of these two lines and surface temperature, metallicity (i.e., Fe abundance), or Ca abundance within the HdC sample, however we do find a relationship between luminosity and the Ca H and K strength. Within our sample, intrinsically brighter stars exhibit weaker Ca II H and K lines. According to \citet{Tisserand2020_plethora} and \citet[][]{Tisserand2022_submitted}, %note was plethora before, not sure on this one.
brighter RCB stars tend to be in the warmer temperature regime. If our analysis was biased to include stars with weak Ca H and K, we would expect most of our Sr-rich class to consist of brighter, and therefore warmer, stars, however we do not see this. In fact, we find no stars in the Sr-rich class that lie in this warmer temperature regime, and most of the Sr-rich stars lie in a narrow range of temperatures at the cooler end of the RCBs and dLHdCs. Thus, we do not suspect that some systematic trend in
Ca H and K strength is affecting our selection of Sr-rich stars.

\begin{figure}
\centering
\includegraphics[width=\columnwidth]{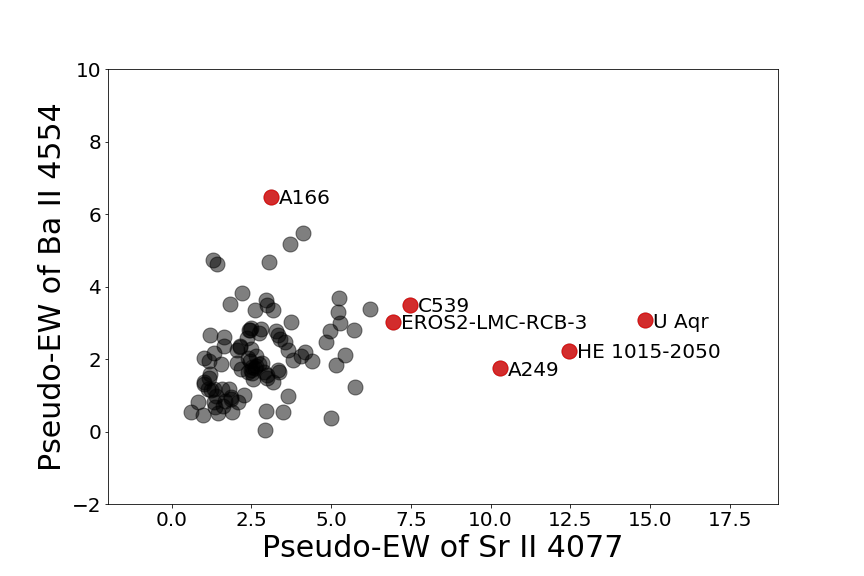}
\caption{The pseudo-EW of Sr II 4077 vs Ba II 4554. The Sr-rich stars are plotted in red and labeled.}
          \label{fig:sr_v_ba}%
\end{figure}

Two other important s-process elements, although less prominent, are observed in the Y II and the Ba II lines. Yttrium is another light \textit{s}-process element, and therefore should be enhanced to a similar extent as the Sr lines are. This line is clearly visible in the spectra of most HdCs, but since it lies in between the Ca II H \& K lines, it is difficult to get an accurate pseudo-EW measurement. Therefore, we inspect this line visually to confirm that the Sr-rich stars are also enhanced in Y II, but otherwise it is not considered. As seen in Figure~\ref{fig:srrichstars}, all Sr-rich stars have clearly identifiable Y II lines except for A166, a unique star which is discussed further in Section~\ref{sec:discussion}.
%We use this spectral line to confirm the Sr in these stars is a result of \textit{s}-processing within the star. 
Barium on the other hand, is a heavy \textit{s}-process element. The strengths of the lines of Sr, Y, and Ba give us an idea on the neutron exposure during nucleosynthesis, as a strong neutron exposure will enhance Ba as well as Sr and Y, whereas a weak exposure will only enhance the lighter elements (see Section~\ref{sec:sprocess}). We observe a large range in the measured ratios of Y/Ba and Sr/Ba pseudo-EWs, see Figure~\ref{fig:sr_v_ba}. The intrinsic spread of Y and Ba abundances has been noted in \citet{Asplund2000} (see Figure 13). They comment that this indicates a large range in the amount of \textit{s}-processing that happens within HdC stars. We note that the star A166 has clearly enhanced Ba as well as Sr and Y, which would signify that its neutron exposure is larger than for U Aqr and the rest of the Sr-rich stars.

HdC stars are generally enhanced in \textit{s}-process elements, as the majority class of RCB stars have average abundances of [Y/Fe] $\simeq$ 0.8 and [Ba/Fe] $\simeq$ 0.4 \citep{Asplund2000}. Notice that this implies the light \textit{s}-process elements are overabundant compared to heavy elements, however not to the same extent as U Aqr which has [Y/Fe] = 3.3 and [Ba/Fe] = 2.1. In our study this translates to HdCs having mostly non-zero pseudo-EWs for the persistent Sr II and Ba II lines, as shown in Figure~\ref{fig:sr_v_ba}.

%describe EW measurements (pseudo-EW)
%list relevant lines (Sr II, Y II, Ba II, Ca H and K)
%plot the stars in order from strongest Sr II lines to weakest
%Use a normal star at similar temperature for comparison
%want to say that dividing by calcium H and K is a valid stand in for dividing by Fe and getting an abundance
%Comment on "abundances" in this low of resolution?

%table of lines used and whether or not they are visible/blended
%consider adding the pseudo-continuum region to the table? Although I don't actually input that directly...

\begin{table}
\caption{Spectral lines relevant to this study}             % title of Table
\label{tab:lines}      % is used to refer this table in the text
\centering                          % used for centering table
\begin{tabular}{c c c c}        
\hline                % inserts double horizontal lines
Element & Wavelength ($\AA$) & Light/Heavy? & Comments \\    % table heading 
\hline                        % inserts single horizontal line
   Sr II & 4077 & Light & \\      % inserting body of the table
   Sr II & 4215 & Light & near C$_2$ bandhead\\
   Ca II & 3933, 3968 &  &  Ca II H \& K \\
   Y II & 3950 & Light & in between H \& K \\
   Ba II & 4554 & Heavy & \\ 
\hline                                   %inserts single line
\end{tabular}
\end{table}

%%%%%%%%%%%%%%%%%%%%%%%%%%%%%%%%%%%%%%%%%%%%%%%%%%%%%%%%%%%%%%%%%%%%%%%%%

%%%%%%%%%%%%%%%%%%%%%%%%%%%%%%%%%%%%%%%%%%%%%%%%%%%%%%%%%%%%%%%%%%%%%%%%%
\section{The \textit{s}-Process in HdC Stars}
\label{sec:sprocess}

Elements heavier than Fe are predominantly synthesized via neutron capture chains in either the slow (\textit{s}-) or rapid (\textit{r}-) processes, the relative contributions of which have been modeled in works such as \citet{Seeger1965_sprocesstheory}. Describing the \textit{s}-process is possible by defining a neutron exposure parameter 
$$\tau [mb^{-1}]=  \int_{0}^{t_{exp}} n_n(t) v_T \,dt  $$
where $n_n(t)$ is the number density of free neutrons, $v_T$ is the thermal velocity, and $t_{exp}$ is the duration of the neutron exposure. There are two types of neutron exposure considered. The first case is a single neutron exposure event and the second is a series of decaying neutron exposure events. The second case corresponds to physical situations such as the thermal pulses in AGB stars where between each pulse the population of neutrons decays. This can be represented mathematically as a weighted exponential function
$C e^{-\tau/\tau_0}$
where C is a weighting constant and $\tau_0$ represents how fast the neutron exposure decays between pulses. When describing a neutron exposure event, $\tau$ is used to describe single neutron exposures and $\tau_0$ is used to describe exponential exposures. A thorough explanation of this theory can be found in works such as \citet{Clayton1961_sprocess}, \citet[ch.~7]{Clayton1968_ch7}, and an exact solution of the exponential case is found in \citet{Clayton1974_exactsoln}.

The measured \textit{s}-process solar system abundances of heavy elements require two components to get a good model fit: the "strong" and the "weak" components \citep{Beer1989_sprocess}. The strong component, also referred to as the "main" component, can be replicated by an exponential neutron exposure with $\tau_0 = (0.30 \pm 0.01)(kT/30eV)^{1/2} [mb^{-1}]$ (Beer 1986c). The weak component, however, is more correctly modeled by a single neutron exposure with $\tau = (0.23 \pm 0.03)(kT/30eV)^{1/2}[mb^{-1}]$ \citep{Beer1989_sprocess}. The strong component enhances all \textit{s}-process abundances, whereas the weak component mainly enhances the light \textit{s}-process elements (i.e., elements with mass number $\leq$ 90). This is because $^{88}$Sr, $^{89}$Y, and $^{90}$Zr all have closed neutron shells and, by extension, small neutron capture cross sections. The weaker neutron exposure associated with the weak component does not provide enough neutron flux for significant neutron capture onto these closed shells, therefore causing a buildup of the lighter elements. 

Previous studies on U Aqr attempted to characterize its neutron exposure event. \citet{Bond1979_uaqr} noted that an enhancement of the light \textit{s}-process elements such as Sr and Y is only possible with weak neutron exposures, and is less likely to occur in an exponential exposure. This is corroborated by \citet{Malaney1987_sprocess} which notes that in single neutron exposures, the Sr/Ba ratio peaks near $\tau$ = 0.5 mb$^{-1}$, with higher exposures having larger amounts of Ba. \citet{Bond1979_uaqr} estimates the best fit for U Aqr as a single neutron exposure with $\tau$ = 0.6 mb$^{-1}$, assuming dilution of the processed material by a factor, f $\approx$ 10. \citet{Malaney1985_uaqr} assumes a post-AGB origin of U Aqr and therefore fits the star's abundances using an exponential neutron exposure with $\tau_0$ = 0.1 mb$^{-1}$. \citet{Vanture1999_uaqr} point out that qualitatively, the \textit{s}-process abundances in U Aqr are similar to those of the solar weak component. They compare their abundance calculations to the general \textit{s}-process models presented in  \citet{Malaney1987_sprocess} and find that it requires a large fraction of processed material in the outer atmosphere to match model abundances. They find the best fit to be a single neutron exposure with $\tau$ = 0.15 to 0.4 mb$^{-1}$, but do not exclude the possibility of an exponential irradiation.

Recently, advances have been made in narrowing down the more likely origin of RCB stars to a double WD merger, and therefore an AGB-like neutron irradiation is unlikely in these kinds of stars. In fact, recent stellar evolution models from \citet{Crawford2020_mesa} and \citet{Munson2021} show that RCB surface abundances are set within the first $\sim$10-50 years of evolution after the WD merger, hundreds of years before reaching the RCB phase, and the surface abundances do not change for the remainder of the star's evolution. This happens because as the star expands due to energy generation at the He-burning shell, the convective region in the outer envelope splits into two regions and the processed material can no longer be transported from the burning region to the surface of the star. Therefore, these models imply that the neutron exposure creating the \textit{s}-process abundances on the surface of RCB stars would resemble a single neutron exposure.

\begin{figure}
\centering
\includegraphics[width=\columnwidth]{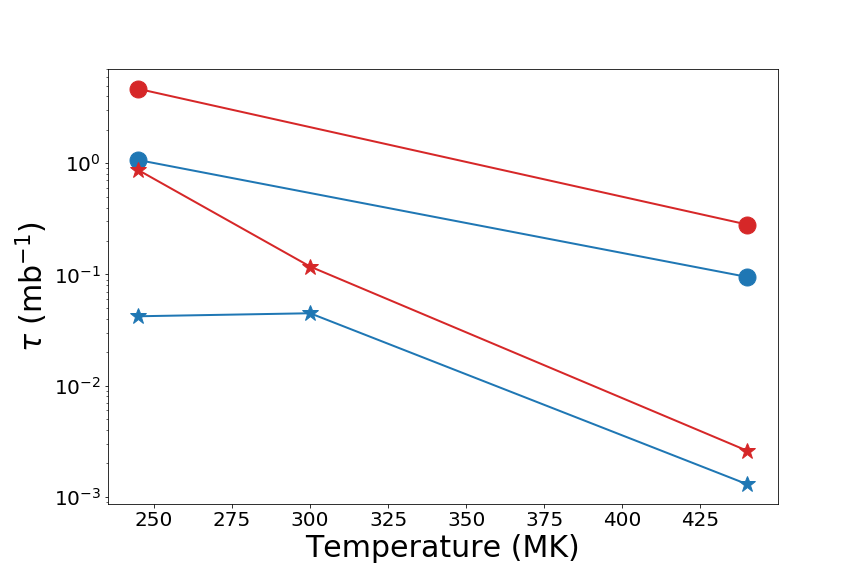}
\caption{Neutron Exposure $\tau$ in mb$^{-1}$ vs RCB model temperature in MK. Red lines indicate a burning region defined as T > 200 MK, blue lines indicate a burning region defined by T > 100 MK. Closed circles denote solar metallicity models and stars indicate subsolar metallicity models. All models detailed in \protect{\citet{Crawford2020_mesa}}.}
          \label{fig:tau}%
\end{figure}

Further, using the models from \citet{Crawford2020_mesa} we can estimate the neutron exposure for a typical HdC star. 
The neutron exposure was calculated for five RCB models: SOL8.39, SOL8.69, SUB8.39, SUB8.48, and SUB8.69. An important aspect of calculating $\tau$ is to define the site where the \textit{s}-process will occur. For this work we defined this region in two ways, first where the temperature of the zone is greater than 10$^{8}$ K (where the triple-$\alpha$ process occurs) and the second where the zone's temperature is greater than 2x10$^{8}$ K (for exploratory purposes). The number density of neutrons and the thermal velocity for each time step was taken as an average over this region. The second important aspect of this calculation is the duration of the neutron exposure, which we took to equal the star's age when the envelope ceased to be fully convective, assuming convective regions to be fully mixed at each time step due to the short dynamical timescale in the models (see Figure 12 in \citet{Crawford2020_mesa} and discussion therein). The resultant values for the neutron irradiation parameter $\tau$ are listed in Table~\ref{tab:tau} and plotted in Figure~\ref{fig:tau}. 

Three trends can be gleaned from these models. First, models with hotter burning regions such as SOL8.69 and SUB8.69 have significantly weaker neutron irradiations than cooler models. The warmer temperature of these models is such that while there is now a second activated neutron source due to the reaction $^{22}$Ne($\alpha$,n)$^{25}$Mg, the neutron poison reaction $^{14}$N(n,p)$^{14}$C is also more strongly activated, which consumes the neutrons quickly, before they have a chance to build up and be captured by iron seed nuclei. 
%The second trend is that choosing a warmer site for neutron capture, and therefore a smaller region, results in a larger neutron exposure. 
Secondly, as each zone averaged under this warmer condition is at higher temperatures and higher neutron densities, the overall average neutron density and temperature over the neutron capture site is larger, resulting in a larger effective neutron exposure. Care must be taken to ensure the correct \textit{s}-process site is chosen. The final trend is that the models with subsolar metallicity have smaller neutron exposures than those with solar metallicity. There are a multitude of reasons this could be, however a strong case could be made by noting that the initial H abundances of these models are not the same. The SOL set of models has an initial mass fraction of H on the order of 10$^{-4}$ for the homogeneous envelope, whereas the SUB set of models has 10$^{-5}$ for the mass fraction of H. It is unclear whether this is the primary cause of the disparity in neutron irradiation, however the increased H in the solar metallicity models would create more neutrons per zone due to the reaction chain $^{12}$C(p,$\gamma$)$^{13}$C($\alpha$,n)$^{16}$O which converts protons into neutrons. Note that this difference in H abundances between the two types of models is not physically motivated, and is instead a result of the process of creating the models. In future models, the H abundance in the envelope will be standardized between different metallicities.

It is clear that the neutron exposures calculated for these models are lower than those estimated for U Aqr, which is understood to have a weaker neutron exposure than normal RCB stars. The lowest irradiation calculated for U Aqr is $\tau$ = 0.15 mb$^{-1}$, whereas the largest irradiation calculated for the preferred RCB Model (SUB8.48) is $\tau$ = 0.1 mb$^{-1}$. Using the $^{56}$Fe neutron capture cross section from \citet{Liou2017_crosssection} of $\sigma$ = 7.5 $\pm$ 4.2 mb at 24.37 keV, a neutron exposure of 0.1 mb$^{-1}$ translates to approximately 0.75 neutrons captured per $^{56}$Fe nucleus. This low level of exposure cannot explain the \textit{s}-process abundances in the majority of RCB stars. Most RCB stars have \textit{s}-process enhancements indicative of stronger neutron exposures than that of U Aqr (i.e. they have smaller Sr/Ba ratios), and our models cannot currently explain this discrepancy. Using a separate test set of unpublished RCB models where the initial envelope mass fraction of N was decreased from 10$^{-2}$ to 10$^{-3}$, we calculated an increase in the neutron irradiation $\tau$ from 0.04 to 8.48. Therefore it is clear that the initial abundance of N plays a large role in the resulting neutron exposure, as it would also govern the presence of the neutron poison, $^{14}$N. More work must be done on interplay of initial N abundance and available neutrons in RCB models.

\begin{table}
\caption{Neutron Exposure ($\tau$) for RCB Models}             % title of Table
\label{tab:tau}      % is used to refer this table in the text
\centering                          % used for centering table
\begin{tabular}{l|c|c}        
RCB Model & 10$^{8}$ K & 2x10$^{8}$ K \\    % table heading 
\hline                        % inserts single horizontal line
   SOL8.39 & 1.1 & 4.7 \\      % inserting body of the table
   SOL8.69 & 0.1 & 0.3 \\
   SUB8.39 & 0.04 & 0.9 \\
   SUB8.48 & 0.05 & 0.1 \\
   SUB8.69 & 0.001 & 0.003 \\ 
\end{tabular}
\end{table}

%%%%%%%%%%%%%%%%%%%%%%%%%%%%%%%%%%%%%%%%%%%%%%%%%%%%%%%%%%%%%%%%%%%%%%%%%

\section{Discussion}
\label{sec:discussion}

Previously, RCB stars and dLHdC stars have been indistinguishable in their abundances derived from optical spectra \citep{Warner1967_hdcs} The only known difference between the two classes is whether or not they form dust. K-band spectra of CO absorption bands has revealed that dLHdC stars generally have larger surface abundance enhancements of $^{18}$O than RCB stars \citep[][]{Clayton2007_O18,GarciaHernandez2009_CNOabunds,Karambelkar2022_submitted}. It is unknown whether or not the dLHdC class is just a dustless phase of RCB evolution. There were previously only five known dLHdC stars, however \citet[][]{Tisserand2022_submitted} has discovered 27 more dLHdC stars, increasing the sample to 32 total stars. Therefore, we can now begin to better discern any further differences between the RCBs and dLHdCs.

Using the spectra of all known RCBs and dLHdCs (Crawford et al. 2022, in prep.), we were able to identify a small sub-class of these stars with extraordinarily enhanced Sr lines, which we denote as Sr-rich stars. The six stars which belong to this sub-class are U Aqr, EROS2-LMC-RCB-3, HE 1015-2050, A249, A166, and C539. The latter four are dLHdCs, the final three of which were recently discovered by \citet[][]{Tisserand2022_submitted}. These stars are easily differentiated from other HdC stars when comparing their Sr II and Ca II H pseudo-EWs (Figure~\ref{fig:sr_v_ca}). There is no clear reason why these six stars have a nearly linear relationship between Sr II and Ca II H pseudo-EWs. These six stars are all in the cooler regime of HdCs (5000-6500K), which clearly show both C$_2$ and CN bands in their spectra. Roughly 2/3 of all HdC stars are within this same temperature regime \citep[][]{Tisserand2020_plethora,Tisserand2022_submitted}, therefore with six Sr-rich stars at cooler temperatures, we would expect to find two warm Sr-rich stars, however no such stars are found. It is not clear whether this kind of Sr enrichment is exclusive to cool stars or if it is a small sample bias. \citet{Jeffery2020_DYCen} points out that the hot RCB star DY Cen shows very strong Sr II enhancement in its 1987 spectrum, indicating that this kind of Sr enhancement can be seen up to surface temperatures of 18800 K. We note for clarity that the hot RCB stars are not included in this study \citep{Demarco2002_hotrcbs,Tisserand2020_plethora}. Compared to the total population of RCBs and dLHdCs, we find that 2 out of the 87 RCBs with blue optical spectra and 4 out of 32 dLHdCs have this kind of Sr enhancement. Therefore, dLHdCs are more likely to be Sr-rich, with an occurrence  rate of $\sim$13\% as opposed to RCBs at $\sim$2\%.

\begin{figure*}
\centering
\includegraphics[width=7in]{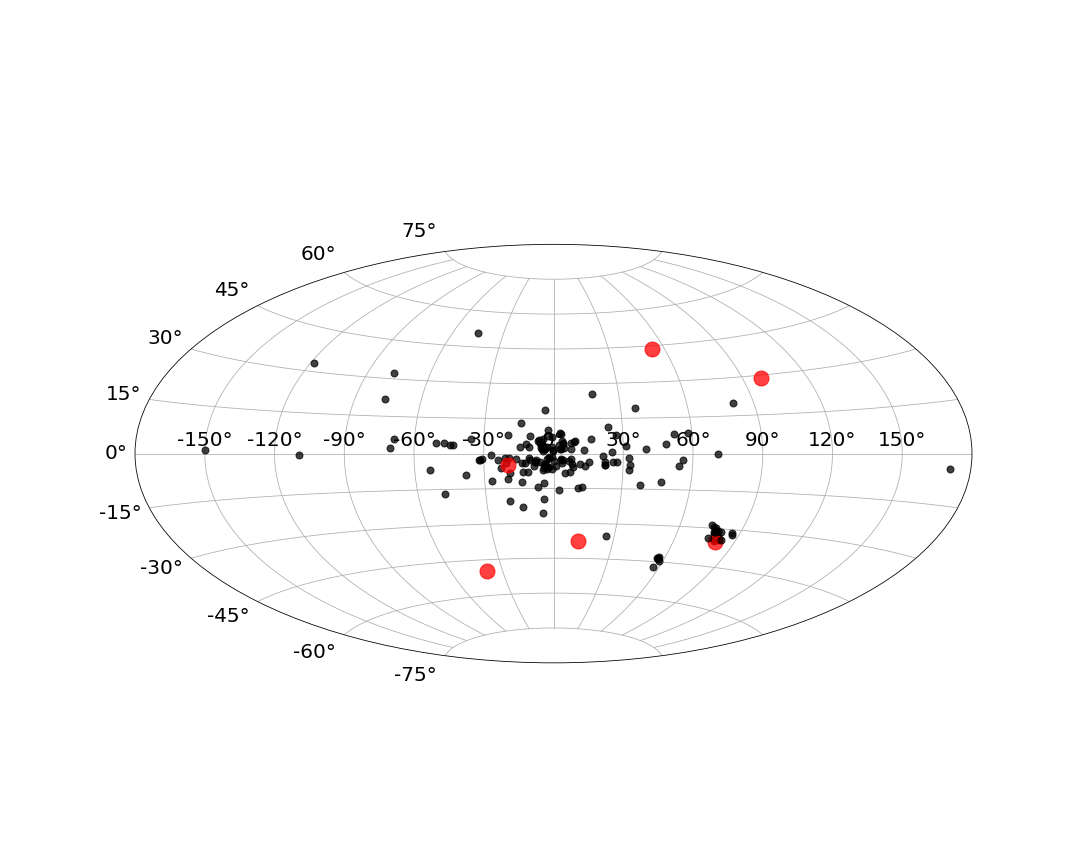}
\caption{The Galactic distribution of all RCBs and dLHdCs, with the Sr-rich stars highlighted in large red dots.}
          \label{fig:galcoords}%
\end{figure*}

We additionally explore the Galactic distribution of the Sr-rich stars, which can be seen in Figure~\ref{fig:galcoords}. Four out of the five Galactic Sr-rich stars lie clearly in the halo, which is known to have very low metallicity. The star that lies in the bulge is C539. We find one Sr-rich RCB in the LMC: EROS2-LMC-RCB-3. We note as well that this star lies away from the majority of the RCBs which lie in the LMC Bar. It seems, then, that this star also lies in the LMC analog to the halo population. There are four other RCBs that lie within the halo population: AO Her, Z Umi, and NSV 11154 in the Northern Galactic Hemisphere, and ASAS-RCB-6 in the Southern Galactic Hemisphere. We do not currently have the spectra necessary to comment on the strength of Sr in these stars. There is one additional star that appears to be in the Northern halo region. This star is the eponymous R CrB, which only lies 1.3 kpc away from us and is therefore not a true member of the halo. The distribution of Sr-rich stars is significantly different from other HdC stars and thus may indicate a correlation between metallicity and the strength of the neutron exposure. 

Recent RCB models such as \citet{Crawford2020_mesa} and \citet{Munson2021} have shown convincing evidence for the origin of the surface abundance of $^{18}$O, a trait unique to the HdC stars. This isotope is created during the reaction chain $^{14}$N($\alpha$,$\gamma$)$^{18}$F($\beta^+$)$^{18}$O($\alpha$,$\gamma$)$^{22}$Ne, which flows to completion in a typical star, causing an enhancement of $^{22}$Ne. However, in these stars, this reaction chain must be halted before completion. These two sets of models show that due to their unusual convective system and its evolution, material is mixed out of the He-burning shell to the surface only within the first $\sim$10-50 years of post-merger evolution, after which the surface abundances are set and do not change throughout the remainder of the star's life. Therefore, this short triple-$\alpha$ burning is integral to the existence of surface $^{18}$O. A comparison can easily be drawn between the this short triple-$\alpha$ burning and the requirement of a weak neutron exposure in RCB stars. This possibly parallel relationship could be further explored using IR spectra of the Sr-rich stars to observe the $^{12}$C$^{18}$O bands. Of the Sr-rich stars, only U Aqr has been observed in the near IR, which confirms it has enhanced $^{18}$O, similar to other RCB stars \citep{Karambelkar2022_submitted}. However, dLHdC stars are known to have larger $^{18}$O enhancements than RCB stars \citep{Clayton2007_O18,GarciaHernandez2009_CNOabunds}, which could explain the larger occurrence rate of Sr-richness in dLHdC stars. More observations are needed to draw any further conclusions on the relationship between surface $^{18}$O and Sr abundances.

The neutron exposure for the Sr-rich class of stars is best understood as a weak, single event. Weaker exposure events can mean either a short time scale, a low neutron density, or a low temperature. As the neutron irradiation $\tau$ relies on the temperature to only the first order, the differences in the He-burning shell temperature from 2.45 to 4.4 x 10$^8$ K are negligible compared to the other factors that influence the neutron exposure. However, we have no evidence to suggest whether the Sr-rich phenomenon is due to short exposure timescales or lower neutron densities. Weak neutron exposure events deposit neutrons onto seed nuclei up until the lowest mass set of closed neutron shells at $^{88}$Sr, $^{89}$Y, and $^{90}$Zr. These nuclei have significantly smaller neutron capture cross sections than the nuclei around them in mass, and therefore the \textit{s}-process slows around these nuclei. This is why Sr-rich stars are associated with weaker neutron exposures, as this buildup of lighter \textit{s}-process elements indicates that neutrons are not captured across closed neutron shells. Upon having stronger exposures, neutron capture across these closed shells becomes more likely, and therefore the \textit{s}-process can progress to create higher mass nuclei such as Ba by capturing onto the lighter nuclei, decreasing their abundance. Thus, as the Ba abundance builds, the lighter elements such as Sr, Y, Zr will decrease in abundance. 

The unique Sr-rich star A166 does not show the typically expected range of Sr and Ba abundances. This star has an enhanced Sr II line as compared to the nearby Ca II H and K lines, however it also shows an enhanced Ba II 4554 line. In Figure~\ref{fig:sr_v_ba}, we can see that this star has a larger pseudo-EW of Ba II 4554 than any of the other HdCs measured. We note that this star is the coldest and reddest known dLHdC, and therefore has quite weak Ca II H\& K, so perhaps the comparison is not straightforward. We also note that the Y II line in this star is weaker than the other Sr-rich stars, perhaps due to the general redness of the spectrum. The enhancement of both Sr II (compared to Ca II H \& K) and Ba II is not something we can easily explain using a single neutron exposure. A166 is unique in almost every way compared to normal dLHdCs \citep{Tisserand2022_submitted,Karambelkar2022_submitted}. The continuum of the spectrum for this star appears different from all other RCBs and dLHdCs except the similarly unique RCB ASAS-RCB-6. The shape of this star's continuum resembles that of a late K star, making it appear extremely red and cold. It also displays an IR excess at 22 $\mu m$~indicating the presence of cold dust. This redness of the spectrum could influence the perceived strength of the lines in the blue and therefore A166 needs to be studied more closely before any conclusions can be drawn regarding its neutron exposure.

There is currently no comprehensive study of \textit{s}-processing in HdCs, so in addition to exploring the sub-class of HdCs with extraordinary Sr enhancement, we also attempt to categorize the neutron exposure in normal HdCs. From the existing estimates of $\tau$ for U Aqr between 0.15 and 0.6 mb$^{-1}$, we estimate that a typical RCB neutron exposure should be greater than 0.6 mb$^{-1}$, however, our calculation using the preferred RCB model from \citet{Crawford2020_mesa} gives $\tau$ = 0.05, assuming the \textit{s}-process occurs at a temperature of 10$^{8}$ K. We are unsure exactly what causes our models to have very small neutron exposures, but we note that increasing the assumed \textit{s}-process temperature to 2x10$^{8}$ increases $\tau$ to 0.1. By investigating the neutron exposure in other RCB models, we find that our subsolar metallicity models have smaller neutron exposures, likely due to the initial abundance of H. Additionally, warmer models have lower neutron exposures due to the larger activation of the $^{14}$N neutron poison. We also found from unpublished models that the initial $^{14}$N abundance plays a large role in the neutron exposure. A lower initial $^{14}$N leads to an increase in the neutron exposure. Further exploration of the neutron exposures in observed HdCs and RCB models are necessary. Note also that our neutron exposure calculations assume a single exposure event rather than pulsed exponential exposures, as the latter of these two options is not reconcilable with our current understanding of RCB formation, especially as our RCB models are of WD merger origin.

In summary, we have found a small sub-class of six RCBs and dLHdCs that are extraordinarily Sr-rich. These stars tend to be in the cooler regime of HdCs, clearly showing both C$_2$ and CN bands, and additionally lie outside of the typical distribution of HdCs in the bulge and old disk regions. One of these stars, A166, is unique even within the class of Sr-rich stars, as it shows strongly enhanced Ba II as well. The expansion of this small class of Sr-rich RCBs and dLHdCs, in tandem with the large increase of known dLHdC stars has allowed us to begin a more detailed exploration into not only the amount of \textit{s}-processing in HdCs, but also the differences and similarities between the two component classes of stars. While dLHdC stars do not exhibit IR dust signatures or the dust declines unique to RCB stars, they do have stronger surface $^{18}$O than RCB stars \citep[][]{Karambelkar2022_submitted} and are more likely to be Sr-rich. \citet[][]{Tisserand2022_submitted} finds that the dLHdCs have stronger H abundances and weaker CN compared to RCBs, as well as occupying a slightly less luminous space of the color-magnitude diagram. These are the only known differences between dLHdCs and RCBs. The Sr-rich class containing both RCBs and dLHdCs provides evidence that these could be different stages of the same stellar evolutionary process, however the spectroscopic differences found in \citet[][]{Karambelkar2022_submitted} and \citet[][]{Tisserand2022_submitted} imply that they may be formed from different WD-binary populations. We are working on obtaining more data to further explore the differences between the two types of HdC stars and discern whether or not they share an evolutionary history.

%%%%%%%%%%%%%%%%%%%%%%%%%%%%%%%%%%%%%%%%%%%%%%%%%%%%%%%%%%%%%%%%%%%%%%%%%

\section*{Acknowlegements}
We would like to thank Catherine Deibel and Scott Marley for their insight on the theory behind the \textit{s}-process. We would additionally like to thank Juhan Frank for his assistance in calculations involving RCB models. CC is grateful for support from National Science Foundation Award 1814967. PT acknowledges also financial support from "Programme National de Physique Stellaire" (PNPS) of CNRS/INSU, France.

This work has made use of data from the European Space Agency (ESA) mission
{\it Gaia} (\url{https://www.cosmos.esa.int/gaia}), processed by the {\it Gaia}
Data Processing and Analysis Consortium (DPAC,
\url{https://www.cosmos.esa.int/web/gaia/dpac/consortium}). Funding for the DPAC
has been provided by national institutions, in particular the institutions
participating in the {\it Gaia} Multilateral Agreement.

%--------------------------------------------------------------------

%-------------------------------------------------------------------

\bibliographystyle{aa}
\bibliography{rcb_bib}

\end{document}